\documentclass[12pt, draftclsnofoot, onecolumn]{IEEEtran}

\usepackage[T1]{fontenc}
\usepackage[cmintegrals]{newtxmath}
\usepackage{times,epsfig,amsfonts,amssymb,url,cite,subfigure,psfrag,booktabs,multirow,bm,rotating,pdfpages,epstopdf,graphicx,gensymb,amsmath}
\begin{document}

\title{Modeling the Multipath Cross-Polarization Ratio for Above-6~GHz Radio Links}

\author{Aki~Karttunen,~\IEEEmembership{Member,~IEEE}, Jan J\"{a}rvel\"{a}inen,~\IEEEmembership{Member,~IEEE}, Sinh Le Hong Nguyen, and Katsuyuki~Haneda,~\IEEEmembership{Member,~IEEE}
	 
\thanks{Manuscript received December 6, 2017. The research leading to these results received funding from the European Commission H2020 programme under grant agreement n$^{\mathrm{o}}$671650 (5G PPP mmMAGIC project), the Academy of Finland and the National Science Foundation through the WiFiUS project "Device-to-Device Communications at Millimeter-Wave Frequencies", the FP7 project ICT-317669 METIS, and the Nokia Bell Labs, Finland.}%
\thanks{A.~Karttunen, S.~L.~H.~Nguyen, and K.~Haneda are with the Aalto University School of Electrical Engineering, 02015 Espoo, Finland. e-mail: \{aki.karttunen, sinh.nguyen, katsuyuki.haneda\}@aalto.fi.}
\thanks{J.~J\"{a}rvel\"{a}inen was with the Aalto University School of Electrical Engineering, Espoo, Finland. He is now with Premix Oy, Finland. e-mail: jan.jarvelainen@premixgroup.com.}
}

\maketitle
\begin{abstract} 
In this paper, we parameterize an excess loss-based multipath component (MPC) cross-polarization ratio (XPR) model in indoor and outdoor environments for above-6~GHz frequency bands. The results are based on 28 measurement campaigns in several frequency bands ranging from 15 to 80~GHz. A conventional XPR model of an MPC assuming a constant mean value fits our measurements very poorly and moreover overestimates the depolarization effect. Our measurements revealed a clear trend that the MPC XPR is inversely proportional to an excess loss in reference to the free-space path loss. The model is physically sound as a higher excess loss is attributed to more lossy interactions or to a greater number of interactions with objects, leading to a greater chance of depolarization. The measurements furthermore showed that the MPC XPR is not strongly frequency or environment dependent. In our MPC XPR model, an MPC with zero-dB excess loss has a mean XPR of 28~dB. The mean XPR decreases half-a-dB as the excess loss increases by every dB and the standard deviation around the mean is 6~dB. The model is applicable to existing channel models to reproduce realistic MPC XPRs for the above 6-GHz radio links. 

\end{abstract}


\IEEEpeerreviewmaketitle


\section{Introduction}
\label{secI_intro}

The above-6 GHz, i.e., centimeter-wave (cm-wave) and millimeter-wave (mm-wave), channels are seen as good candidates for achieving ultra-high throughput for the next generation of wireless communications. 
The main advantage of the cm- and mm-wave channels, compared to the sub-6~GHz, is the availability of greater bandwidth. 
One of the propagation channel properties that affect the communication systems is the radio propagation channel polarization properties. 
The polarization properties determine, e.g., the usefulness of polarization multiplexing, diversity, or severity of polarization mismatch. 
The last is particularly significant at above-6~GHz band~\cite{Maltsev2010, Dupleich2016}. 

The wave depolarization is typically characterized by the cross-polarization ratio (XPR) for each multipath component (MPC). 
It is the ratio of propagation path attenuation when a wave is transmitted and received with the same polarization to when a wave is transmitted and received at the orthogonal polarization. 
It is worth mentioning that the XPR ideally does not depend on the ability of the transmit (Tx) and the receive (Rx) antennas in cross-polar discrimination, and that the XPR denotes depolarization during wave propagation between the antennas.

Cross-polarization ratio measurements that have been carried out in the above-6 GHz bands show XPR values in the range of 10~--~30~dB \cite{Karttunen2017, Karttunen2015, MacCartneyIEEEAccess15, GuerinVTC96, KyroEuCAP10, MaltsevJSAC09, Roivainen2016, Kim2014, Gustafson2016, Vehmas2016}. 
However, a part of the papers~\cite{MaltsevJSAC09, Roivainen2016, Kim2014, Zhu2015} only report the XPR of path loss, which is the ratio of the total power in the main and the cross-polarization observed at the output of the receive antenna ports. 
The path loss XPR changes for different antenna installation at the Tx and Rx and is not a widely valid parameter when it comes to radio propagation modeling. 
We, therefore, focus on modeling the MPC XPR. 
Typically the MPC XPR modeling is adapted in geometry-based stochastic channel models (GSCMs), e.g.,~\cite{3GPP_TR38901, winner}. 
The XPR is defined for cluster sub-paths and is conventionally a log-normally distributed random variable with scenario dependent mean and standard deviation values.

In~\cite{Karttunen2017}, an improved and physically more sound XPR model was proposed, where the mean MPC XPR decreases as an excess loss of the MPC increases. 
Here, the excess loss is a difference between the path power and the free-space path loss corresponding to the path delay. 
It is a more physically sound model as greater excess loss implies more lossy interactions or a greater number of interactions with physical objects, making the depolarization more frequent. 
It is shown that the new model fits the measured MPC XPR better than the conventional model with a constant mean value in an indoor environment at 60-GHz radio frequency~\cite{Karttunen2017}. 
In this paper we extend the work~\cite{Karttunen2017} in threefold: 
\begin{itemize}
	\item The MPC XPR is studied in many different indoor and outdoor environments based on a total of 28 measurement campaigns
	\item The frequency dependency of the MPC XPR is investigated by parameterizing the excess-loss based model at the cm-wave (15-GHz, 28-GHz), and mm-wave (60-GHz, 70-GHz, and 80-GHz) bands\footnote{Including center frequencies 14.25~GHz, 15~GHz, 27.45~GHz, 28.5~GHz, 61~GHz, 63~GHz, 71.5~GHz, and 83.5~GHz.}
	\item A method for comparing the accuracy of MPC XPR models is established. The method is based on the prediction accuracy of the total cross-polarization power and thus provides a single-number comparison metric that is different than the per-MPC XPR data used in the model parametrization. 
\end{itemize}

To the best of the authors' knowledge, the presented work is the first extensive investigation of the MPC XPR with a wide range of environments and frequencies in the cm- and mm-wave bands. 

The paper is organized as follows. 
Channel depolarization and polarization modeling are discussed in Section~\ref{secII_Modeling_Depolarization}. 
The channel measurements are introduced in Section~\ref{secIII_Measurements}. 
The parameter estimation method are presented in Section~\ref{secIV:parameterestimation}. 
In Section~\ref{secV:XPRmodelComparison}, the MPC XPR models are compared based on the prediction accuracy of total cross-polarization power. 
The frequency and environment dependency of the model are analyzed in Section~\ref{secVI:XPRf}, resulting in the best-fit MPC XPR model and parameters. 
Conclusions are presented in Section~\ref{conclusion}.

\section{Depolarization and Channel Polarization Modeling}
\label{secII_Modeling_Depolarization}

When radio waves propagate from the transmitter to the receiver, they reflect, scatter, and diffract from surrounding objects. 
If the surface is rough or has discontinuities such as edges, the wave will be depolarized, and a part of the power is coupled to the orthogonal polarization~\cite{Bertoni2000}. 
Waves might encounter depolarization also while reflecting from large and smooth surfaces if the surface is tilted with respect to the incidence plane~\cite{Priebe2011}.
In general, when the waves reflect, scatter, or diffract they lose power, the polarization ellipse turns, eccentricity changes and the phase difference between polarization components change, converting the polarization from linear to elliptical. 
Models for the depolarization effects have been developed in~\cite{Yin2015, Jaeckel2012}. 
It should be noted that XPR only describes the power ratio between orthogonal polarizations and does not indicate the phase difference between the orthogonal components or the handedness of the elliptical polarization. 

In the existing GSCMs, e.g.,~\cite{3GPP_TR38901,winner}, the polarization is modeled with a 2~$\times$~2 matrix
\begin{eqnarray}
	\label{M}
	\bm{M}_{n,m} & = &
	\begin{bmatrix} a^{VV} & a^{VH}  \\ a^{HV} & a^{HH} \end{bmatrix} \\
	& = & \begin{bmatrix} e^{j\Phi_{n,m}^{VV}} & \sqrt{{\kappa_{n,m}}^{-1}}e^{j\Phi_{n,m}^{VH}}  \\ \sqrt{{\kappa_{n,m}}^{-1}}e^{j\Phi_{n,m}^{HV}} & e^{j\Phi_{n,m}^{HH}} \end{bmatrix},
\end{eqnarray}
where $\kappa_{n,m}$ is the cross-polarization ratio, in linear scale, of an $m$-th MPC of cluster $n$ and $\Phi_{n,m}^{\alpha_{\rm p}\beta_{\rm p}}$ are random initial phases for the $\alpha_{\rm p}$-Rx and $\beta_{\rm p}$-Tx fields; $\alpha_{\rm p}, \beta_{\rm p}$ are either vertical $V$ or horizontal $H$ polarizations. 
The same XPR applies to $\left|a_{VV}/a_{HV}\right|^2$ and $\left|a_{HH}/a_{VH}\right|^2$~\cite{3GPP_TR38901, winner}. 
Importantly, the depolarization during wave propagation and the antenna polarization discrimination are modeled separately. The cross-polarization of propagation channels are defined by the radiated and received fields of an MPC. They are independent of the antenna cross-polarization discrimination (XPD) and antenna orientation. 
The XPR of each MPC, i.e., cluster sub-path, has been conventionally modeled with the log-normal distribution as
\begin{equation}
	\mathrm{XPR}|_{\mathrm{dB}} \sim \mathcal{N}(\mu_1,\sigma_1^2),
	\label{eq:XPR1}
\end{equation}
where $\mu_1$ and $\sigma_1$ are the scenario specific mean and standard deviation, respectively. While it is simple to assume that all MPC follows the same XPR statistics, there is no propagation physics that supports the model.

In \cite{Karttunen2017}, an improved and physically more sound model of an MPC XPR was proposed, in which the mean of the log-normal distribution decreases linearly as a function of the MPC excess loss $L_\mathrm{ex}$, in dBs, as
\begin{equation}
\mathrm{XPR}|_{\mathrm{dB}} \sim \mathcal{N}(\mu_2(L_\mathrm{ex}),\sigma_2^2),
\label{eq:XPR2}
\end{equation}
\begin{equation}
\mu_2(L_\mathrm{ex}) = \begin{cases} \alpha_2\cdot L_\mathrm{ex}+\beta_2, & \mbox{if } L_\mathrm{ex} \leq -\beta_2/\alpha_2 \\ 0, & \mbox{if } L_\mathrm{ex} > -\beta_2/\alpha_2, \end{cases}
\label{eq:mu2}
\end{equation}
where $\mu_2(L_\mathrm{ex})$ is the mean, $\sigma_2^2$ is the variance. 
The excess loss $L_\mathrm{ex}$ of the MPC is defined as $L_\mathrm{ex}=P^\mathrm{m}-\mathrm{FSPL}(\tau)$, where $P^\mathrm{m}$ is the main polarization amplitude and $\mathrm{FSPL}(\tau)$ is the free space path loss at the delay $\tau$. 
In this model, the standard deviation $\sigma_2$ is a constant.

The excess-loss dependent model~\eqref{eq:mu2} has two parts and both have a simple intuitive explanation that may help understand the physics behind the model. 
The first part, for $L_\mathrm{ex} \leq -\beta_2/\alpha_2$, states that for a low-loss paths the mean XPR is proportional to the excess loss. 
This is quite logical since both excess loss and depolarization are caused by interactions with the environment, and therefore, MPCs with higher loss are more likely depolarized, e.g., due to multiple reflections. 
Since a low excess loss implies a high XPR, it is fair to assume that the transmitted linear polarization is almost kept and received linear. 
In this case, the mean XPR can be approximated with a mean polarization rotation angle $\gamma$ over different MPCs showing similar excess losses. 
The XPR of a linear polarization depends on polarization rotation angle $\gamma$ as \cite{Jaeckel2012}
\begin{equation}
\label{eq:xprphi1}
\mathrm{XPR}|_{\mathrm{dB}} = 20\cdot \log_{10}\left(1/\tan\gamma\right).
\end{equation}
With large XPRs we can simplify $\tan\gamma\approx\gamma$ and therefore
\begin{equation}
\label{eq:xprphi2}
\mathrm{XPR}|_{\mathrm{dB}} \approx 20\cdot \log_{10}\left(1/\gamma\right).
\end{equation}
Combining \eqref{eq:mu2}, for small $L_\mathrm{ex}$, and \eqref{eq:xprphi2}, we get an expression of the mean polarization rotation angle for MPCs with low excess loss in dB's as, 
\begin{equation}
\label{eq:xprphimodel2}
\gamma = 10^{-\beta/20}\left( 10^{L_\mathrm{ex}/10}\right)^{-\alpha/2}.
\end{equation}
Equation \eqref{eq:xprphimodel2} states that \eqref{eq:mu2}, for small $L_\mathrm{ex}$, can be approximated by a polarization rotation angle proportional to the excess loss (and the XPR model parameters $\alpha$ and $\beta$). 
The second part of \eqref{eq:mu2}, for $L_\mathrm{ex} > -\beta_2/\alpha_2$, describes the mean XPR for very weak MPCs. 
The mean XPR being zero corresponds to elliptical polarization with random $\gamma$ and random eccentricity, leading to a random received polarization. 
In this manner, the simple XPR model dependent on the MPC excess-loss reflects the propagation physics. 

In the following, the conventional and excess loss dependent models, i.e., \eqref{eq:XPR1} and \eqref{eq:XPR2}-\eqref{eq:mu2} are called model~1 and 2, respectively.

\section{Channel Measurements and Path Detection}
\label{secIII_Measurements}

Channel measurements were conducted in different indoor and outdoor sites in the cm-wave (15-GHz, 28-GHz) and mm-wave (60-GHz, 70-GHz, and 80-GHz) bands. 
The measurement campaigns cover a good balance between indoor and outdoor as well as between cm-wave and mm-wave including 28 campaigns (6 indoor cm-wave, 9 indoor mm-wave, 9 outdoor cm-wave, and 4 outdoor mm-wave), 11 different locations (6 indoor and 5 outdoor), and a total of 265 measured links (139 indoor and 126 outdoor).

\subsection{Environments under Study}

General descriptions of the environments are given below. 
A summary of the frequencies, bandwidths, antenna heights, and the number of links is given in Table~\ref{table:Numbers}. 
These measurements usually cover links with one base station (BS) location and several mobile station (MS) locations. 
A photograph, or a reference to a photograph, is provided for each measurement site and the center frequencies of the campaigns are mentioned.

\subsubsection{Shopping Mall (SHOP)}
The shopping mall Sello is a modern four-story building in Espoo, Finland, with long corridors and a large open space in the middle. The dimensions are roughly 120$\times$70 m$^2$. 
A photo of the shopping mall is shown in~\cite{Karttunen2015VTC}\footnote{Measurements reported in~\cite{Karttunen2015VTC} are different than the ones analyzed in this paper but the shopping mall is the same.}.
Measurements included four center frequencies: 15~GHz, 28.5~GHz, 63~GHz~\cite{Haneda2016}, and also at 71.5~GHz~\cite{Haneda2014,Haneda2015,Karttunen2015}. 

\subsubsection{Airport (AIR)}
The airport measurements were conducted in the check-in area of terminal 2 of Helsinki airport, summarized in~\cite{Vehmas2016}. 
The BS was overlooking the check-in area, while the MS were deployed in the main hall and in a connecting corridor, leading to line-of-sight (LOS) and non-LOS (NLOS) links, respectively. 
The measurement campaigns at center frequencies 15~GHz, 28.5~GHz, and 61~GHz are described in~\cite{Vehmas2016}. 
Also, measurements at 83.5~GHz are included, mostly with the same Tx and Rx antenna locations. 

\subsubsection{Cafeteria (C1)}
Measurements were conducted in a coffee room shown in~\cite{Virk2017}. 
Measurements include both links within the coffee room and room-to-corridor links, leading to LOS and NLOS channels, respectively. 
Three center frequencies were covered: 15~GHz, 28.5~GHz, and 61~GHz~\cite{Virk2017}.

\subsubsection{Cafeteria (C2)}
Measurements in indoor cafeteria room were conducted at 63~GHz center frequency~\cite{Karttunen2017, Jarvelainen2016}.
The cafeteria room is about  14~$\times$~13.5~$\times$~2.8~m$^3$.

\subsubsection{Empty Office (OFF1) and Office in Use (OFF2)}
\begin{figure}[!t]
	\centering
	\includegraphics[trim=0mm 0mm 0mm 0mm,width=50mm]{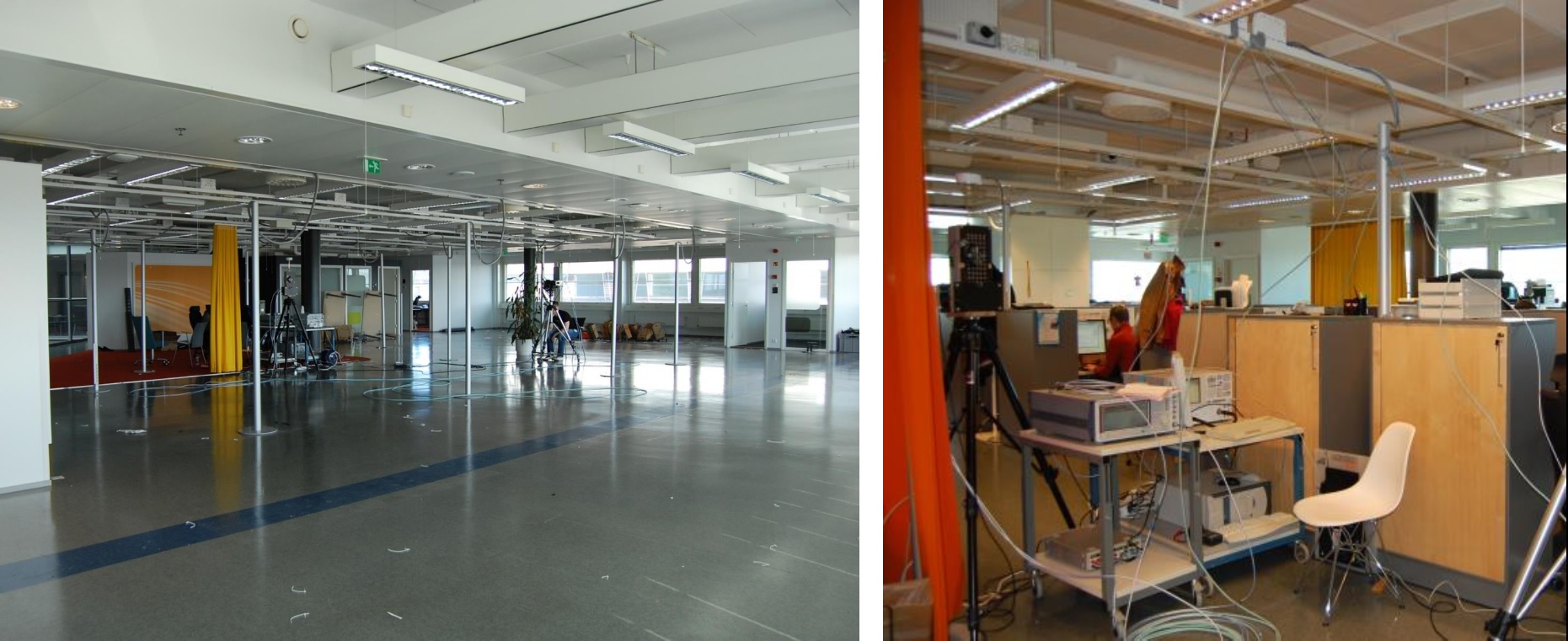}
	\caption{Photos of the empty office (OFF1) and office in use (OFF2) measurement sites.}
	\label{OFF1_OFF}
\end{figure}

Measurements in an office at 71.5~GHz are described in \cite{Haneda2014,Jarvelainen2014,Haneda2015,Karttunen2015}. 
Channel measurements covered similar office rooms with and without the furniture, i.e., empty office (OFF1) and office in use (OFF2), as shown in Fig.~\ref{OFF1_OFF}. 
The office is in a modern building, with a dimension of 18~$\times$~22~$\times$~2.5~m$^3$.


\begin{table*}[t!]
	\caption{Parameters for 28 Measurement Campaigns: Center Frequency ($f$), Bandwidth (BW), Antenna Heights ($h_{\mathrm{BS}}$, $h_{\mathrm{MS}}$), Link Distance Range ($d_{\mathrm{min}}$--$d_{\mathrm{max}}$), Dynamic Range ($\max(P_i^\mathrm{m})-P_{\mathrm{th}}$), Number of Links (Links), Number of MPC with Measured XPR ($XPR=P_i^\mathrm{m}-P_i^\mathrm{c}$), Censored Cross-Polarization ($XPR>P_i^\mathrm{m}-P_{\mathrm{th}}$), and Censored Main Polarization ($XPR<P_{\mathrm{th}}-P_i^\mathrm{c}$), the MPC XPR Model Parameter Estimates ($\hat{\mu}_1$, $\hat{\sigma}_1$, $\hat{\alpha}_2$, $\hat{\beta}_2$, $\hat{\sigma}_2$) and the Total Cross-Polarization Estimation Accuracy ($\mu_\epsilon$).}
	\label{table:Numbers}
	\centering
	\resizebox{\textwidth}{!}{%
		\begin{tabular}{cc||cccc|cccc|ccc|c|cc|c}
			\multicolumn{2}{c||}{} & {\bf SHOP} & {\bf SHOP} & {\bf SHOP} & {\bf SHOP} & {\bf AIR}  & {\bf AIR}  & {\bf AIR}  & {\bf AIR} & {\bf C1}& {\bf C1}& {\bf C1}& {\bf C2} & {\bf OFF1}& {\bf OFF2}& {\bf STA}\\
			\hline 
			\hline
			\multicolumn{2}{c||}{$f$ [GHz]} & 15 & 28.5 & 63 &71.5 & 15 & 28.5 & 61 & 83.5 & 15 & 28.5 & 61& 63 &71.5&71.5&71.5\\
			\multicolumn{2}{c||}{BW [GHz]} &2&3&4&5 &2&3&4&4&2&3&4&4&5&5&5\\
			\hline
			\multicolumn{2}{c||}{$h_{\mathrm{BS}}$ [m]} &1.9&1.9&1.9&1.9 & 5.7&5.7&5.7&5.7& 1.9&1.9&1.9& 2& 1.9&1.9&1.9 \\
			\multicolumn{2}{c||}{$h_{\mathrm{MS}}$ [m]} &1.9&1.9&1.9&1.9 & 1.6&1.6&1.6&1.6& 1.9&1.9&1.9& 2& 1.9&1.9&1.9\\
			\multicolumn{2}{c||}{$d_{\mathrm{min}}$--$d_{\mathrm{max}}$ [m]} &5--65&5--48&5--48&1-9 &15--107&15--90&18--83&18--83&4--7&4--12&3--7&3--7 & 2--10 & 1--8& 1--6\\
			\multicolumn{2}{c||}{$\max(P_i^\mathrm{m})-P_{\mathrm{th}}$ [dB]}  &36&38&43&35&33&32&35&39&43&42&51&54&34&41&44 \\
			\multicolumn{2}{c||}{Links} &9&9&7&17&9&10&8&8&15&7&4&7&12&9&8 \\
			\hline
			\hline
			\multicolumn{2}{c||}{$XPR=P_i^\mathrm{m}-P_i^\mathrm{c}$}           &144&265&219&22&153&132&207&404&218&73&355&145&76&54&34 \\
			\multicolumn{2}{c||}{$XPR>P_i^\mathrm{m}-P_{\mathrm{th}}$} &885&978&983&866&612&348&753&1070&687&172&374&936&1587&1054&444 \\
			\multicolumn{2}{c||}{$XPR<P_{\mathrm{th}}-P_i^\mathrm{c}$} &8&24&34&0&22&33&85&90&13&7&34&2&5&2&5 \\
			\hline
			\hline
			\rule{0pt}{3ex}    
			\multirow{3}{*}{\begin{turn}{90}\bf {\scriptsize Model 1   }\end{turn}}\kern-5mm&$\hat{\mu}_1$ &18&14&18&34&12&11&12&12&18&14&14&22&26&26&29 \\
			&$\hat{\sigma}_1$ &8.8&8.2&11&12&8.1&9.0&11&9.6&8.5&8.5&11&9.4&9.4&9.5&12 \\
			&$\mu_\epsilon$  &4&4&8&4&5&8&12&10&4&6&12&7&5&5&7 \\
			\hline
			\hline
			\rule{0pt}{2ex}  
			\multirow{4}{*}{\begin{turn}{90}\bf {\scriptsize Model 2   }\end{turn}}\kern-5mm&$\hat{\alpha}_2$  &-0.55&-0.57&-0.77&-0.69&-0.59&-0.70&-0.76&-0.66&-0.46&-0.46&-0.60&-0.61&-0.55&-0.49&-0.53 \\
			&$\hat{\beta}_2$   &29&30&36&42&24&26&31&29&28&25&31&36&34&33&39 \\
			&$\hat{\sigma}_2$  &6.2&6.0&6.9&6.9&6.1&6.5&8.4&7.2&5.9&5.8&8.1&3.6&5.4&6.1&8.1 \\
			&$\mu_\epsilon$  &-1&-1&0&1&1&3&5&3&0&3&1&-1&0&1&2 \\
			\hline
	\end{tabular}}
	\begin{tabular}{c}
		\\
	\end{tabular}
	\centering
	\resizebox{\textwidth}{!}{%
		\begin{tabular}{cc||ccccc|cc|c|cccc|c}
			\multicolumn{2}{c||}{} & \bf SQR1 & \bf SQR1 & \bf SQR2 & \bf SQR2 & \bf SQR2  & \bf SC1 & \bf SC1  & \bf SC2 & \bf SC3 & \bf SC3 & \bf SC3 & \bf SC3 & \bf SC4 \\  
			\hline
			\hline
			\multicolumn{2}{c||}{$f$ [GHz]}& 27.45&83.5&14.25&27.45& 61 &14.25&27.45&27.45&14.25 &27.45 & 61 & 83.5 &27.45\\ 
			\multicolumn{2}{c||}{BW [GHz]} &0.9&4&0.5&0.9&4&0.5&0.9&0.9&0.5&0.9&4&4&0.9\\
			\hline
			\multicolumn{2}{c||}{$h_{\mathrm{BS}}$ [m]}&5&5&    2.6&2.6&2.6&2.6&2.6&2.6&2.6&2.6&2.6&2.6&2.6\\
			\multicolumn{2}{c||}{$h_{\mathrm{MS}}$ [m]}&1.6&1.6&2.6&2.6&2.6&2.6&2.6&2.6&2.6&2.6&2.6&2.6&2.6\\
			\multicolumn{2}{c||}{$d_{\mathrm{min}}$--$d_{\mathrm{max}}$ [m]}&10--61&29--61&5--99&5--99&5--65&10--115&10--148&10--188&16--121&16--121&35 --120&16--94&15--90 \\
			\multicolumn{2}{c||}{$\max(P_i^\mathrm{m})-P_{\mathrm{th}}$ [dB]} &38&37&42&43&44&42&45&47&43&45&50&44&38 \\
			\multicolumn{2}{c||}{Links} &7&6&10&12&10&11&12&13&11&11&7&5&11 \\		
			\hline
			\hline
			\multicolumn{2}{c||}{$XPR=P_i^\mathrm{m}-P_i^\mathrm{c}$} &68&114&464&290&422&154&129&335&42&232&322&290&248 \\   
			\multicolumn{2}{c||}{$XPR>P_i^\mathrm{m}-P_{\mathrm{th}}$} &638&1483&1399&1571&2439&583&667&1441&176&472&781&579&696 \\
			\multicolumn{2}{c||}{$XPR<P_{\mathrm{th}}-P_i^\mathrm{c}$} &7&1&10&15&48&3&7&12&3&15&27&27&21 \\		\hline
			\hline
			\rule{0pt}{3ex}    
			\multirow{3}{*}{\begin{turn}{90}\bf {\scriptsize Model 1}\end{turn}}\kern-5mm&$\hat{\mu}_1$ &19&21&16&16&18&15&15&16&15&12&14&12&13 \\
			&$\hat{\sigma}_1$ &8.7&6.8&6.9&7.5&10&6.4&7.2&6.6&6.8&7.3&8.7&8.3&7.7 \\
			&$\mu_\epsilon$ &5&4&4&5&6&4&4&8&10&4&8&6&2 \\
			\hline
			\hline
			\rule{0pt}{2ex}  
			\multirow{4}{*}{\begin{turn}{90}\bf {\scriptsize Model 2}\end{turn}}\kern-5mm&$\hat{\alpha}_2$ &-0.56&-0.47&-0.44&-0.47&-0.57&-0.34&-0.48&-0.47&-0.37&-0.31&-0.31&-0.34&-0.35 \\
			&$\hat{\beta}_2$  &30&29&27&27&33&23&28&27&22&21&23&23&23 \\
			&$\hat{\sigma}_2$ &5.3&4.3&5.0&5.4&6.8&5.2&5.2&4.7&5.0&6.1&7.5&7.0&6.5 \\
			&$\mu_\epsilon$ &0&1&-1&0&0&3&5&2&7&1&4&0&-1 \\
			\hline
	\end{tabular}}
\end{table*} 


\subsubsection{Railway Station (STA)}
\begin{figure}[!t] 
	\centering
	\includegraphics[trim=0mm 0mm 0mm 0mm,width=50mm]{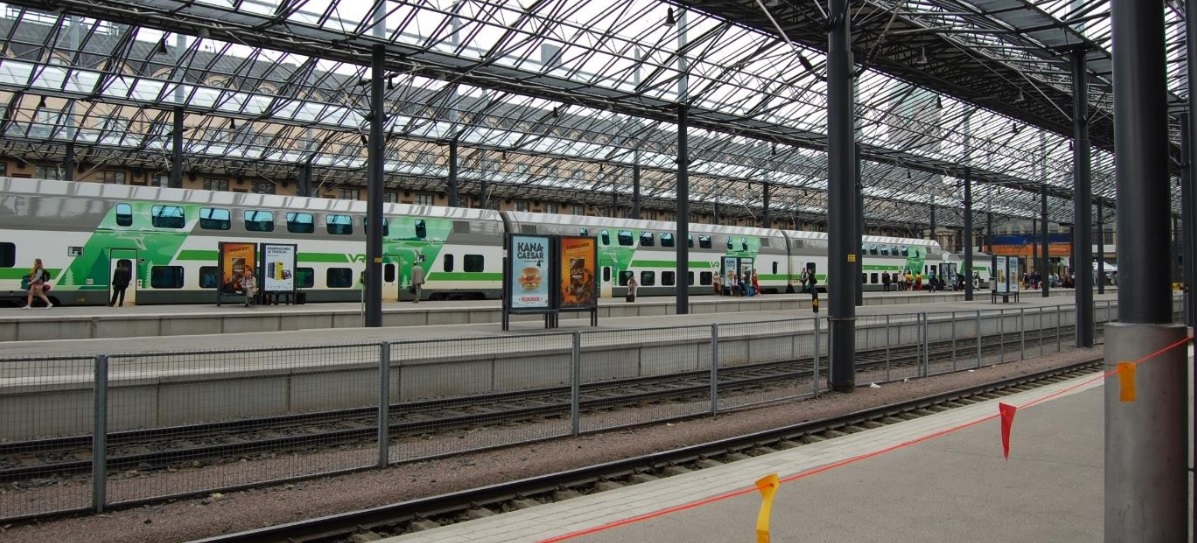}
	\caption{Photo of the railway station (STA) measurement site.}
	\label{STA}
\end{figure}

Measurements in Helsinki central railway station platform at 71.5~GHz are described in~\cite{Haneda2014, Haneda2015, Karttunen2015}. 
A photograph of the site is shown in Fig.~\ref{STA}. 

\subsubsection{Open Square (SQR1 and SQR2)}
The open square Narikkatori, presented in~\cite{Nguyen2015}\footnote{Measurements used in~\cite{Nguyen2015} are different than the ones analyzed in this paper but the location is the same.}, is located in Helsinki, Finland. 
It has approximate dimensions of 90~$\times$~90~m$^2$ and is surrounded by modern, multi-story buildings. 
The environment does also contain lamp posts, trees, and a sculpture. 
Measurements have been performed with different antenna heights: $h_{\mathrm{BS}}=$~5~m and $h_{\mathrm{MS}}=$~1.6~m for 27.45~GHz and 84.5~GHz (SQR1), $h_{\mathrm{BS}}=h_{\mathrm{MS}}=$~2.6~m for 27.45~GHz and 61~GHz (SQR2).

\begin{figure}[!t]
	\centering
	\subfigure[]{
		\includegraphics[trim=0mm 0mm 0mm 0mm,width=50mm]{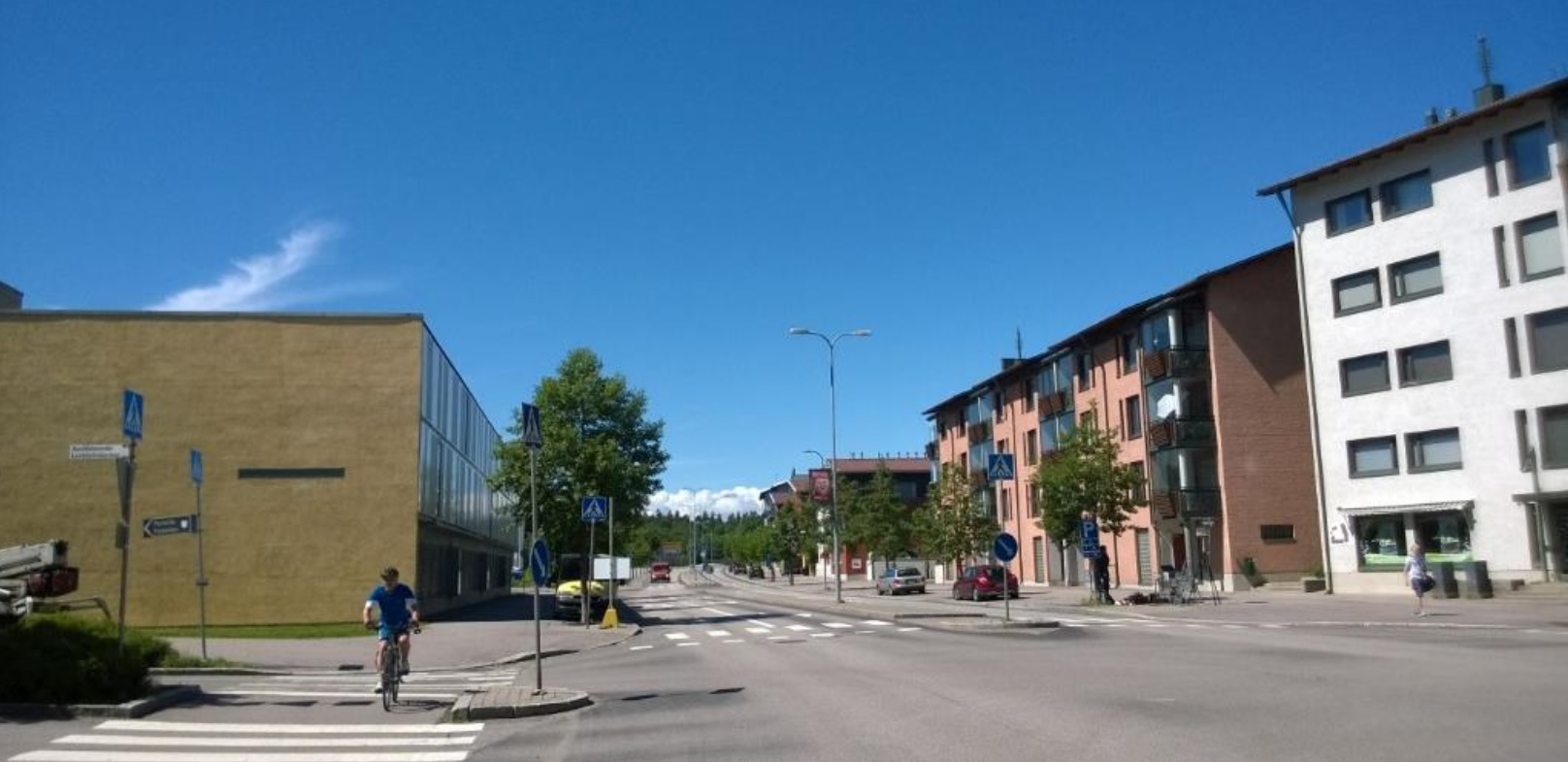}
		\label{fig:photo_SC1}}
	\subfigure[]{
		\includegraphics[trim=0mm 0mm 0mm 0mm,width=50mm]{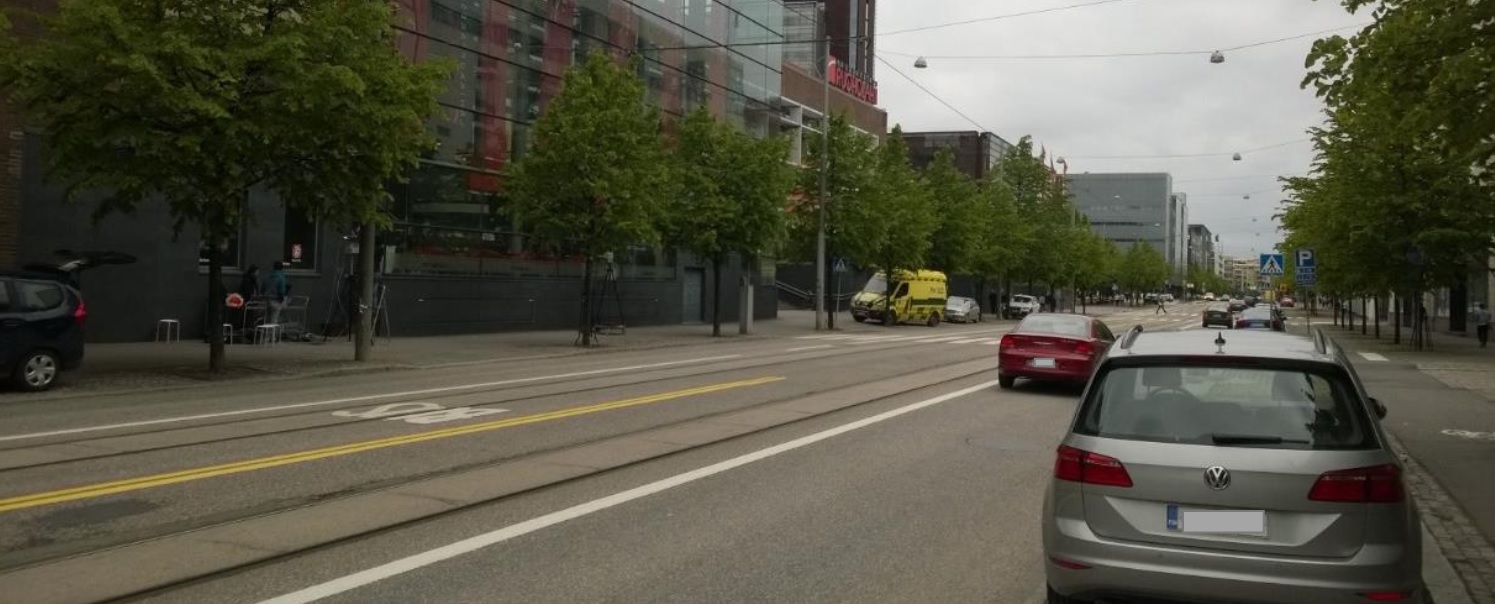}
		\label{fig:photo_SC4}}
	\caption{Photos of the street canyons (a) SC1 and (b) SC4.}
\end{figure}

\subsubsection{Street Canyons (SC1 -- SC4)}
Street canyon measurements are conducted in four different streets. 
The first street (SC1), called Lepp\"avaarankatu, is located in a residential area in Espoo, Finland. 
The street is roughly 30 m wide, and has trees and balconies along the buildings, as depicted by Fig.~\ref{fig:photo_SC1}.
Measurements have been made with center frequencies 14.25~GHz and 27.45~GHz. 
The second street (SC2), Aleksanterinkatu, is located in downtown Helsinki, Finland. 
A photo of the street is shown in~\cite{Nguyen2016Globecom}.
Measurements have covered center frequency of 27.45~GHz~\cite{Nguyen2016Globecom}. 
The third street (SC3) is located in Espoo, Finland, at Aalto University campus. 
A photo of the street is shown in~\cite{Naderpour2016}. 
The measurement campaigns at center frequencies 14.25~GHz, 27.45~GHz, and 61~GHz are described in~\cite{Naderpour2016}.  
Also, measurements at 83.5~GHz are included with the same Tx and Rx antenna locations. 
The fourth street (SC4), It\"amerenkatu, is located in downtown Helsinki, Finland. A photo of the street is shown in Fig.~\ref{fig:photo_SC4}. 
Measurements have been performed with the center frequency of 27.45~GHz.


\subsection{Channel Sounders}

The 28 measurement campaigns were conducted by Aalto University between 2013 and 2016. 
The channel sounder setups can be divided into two groups: the 60- and 70-GHz band sounder used and described in~\cite{Karttunen2017, Jarvelainen2014, Jarvelainen2016, Haneda2014, Haneda2015, Karttunen2015, Nguyen2015, Karttunen2015VTC} and the long-range multi-frequency sounder setup in~\cite{Naderpour2016, Nguyen2016Globecom, Vehmas2016, Virk2017}.
A brief description of the sounders and used antennas is given in the following.

\subsubsection{60- and 70-GHz Band Sounder}

A vector network analyzer (VNA)-based channel sounder is used with up and down converters that cover both the 60- and 70-GHz bands~\cite{Karttunen2017, Jarvelainen2014, Jarvelainen2016, Haneda2014, Haneda2015, Karttunen2015, Nguyen2015, Karttunen2015VTC}.
Five out of the 28 campaigns are measured with this sounder setup, i.e., SHOP and C2 at 63~GHz and OFF1, OFF2, and STA at the 71.5-GHz center frequency. 
Both the Tx and the Rx side are connected to the VNA and to the signal generator, providing the local oscillator signal, by coaxial cables, limiting the link distance range to about 20~m.\footnote{The optical fiber cables are used in SHOP at 63~GHz with the 60- and 70-GHz band sounder-setup.} 
On the Rx side, there is an omnidirectional antenna with 5~dBi maximum gain and 11\degree{} elevation beamwidth. 
The Tx antenna is a standard gain horn with 20~dBi gain with about 20\degree{}  beamwidth in both azimuth and elevation. 
The measurements covered Tx- and Rx-vertical (VV)\footnote{Except in one measured link in C2 at 63~GHz, where the main polarization is horizontal-to-horizontal (HH) and the cross-polarization measurement was vertical-to-horizontal (VH). In this measurement the horn is scanned in elevation plane as explained in \cite{Karttunen2017}.} 
and Tx-horizontal/Rx-vertical (VH) measurements. 
In the latter case, the horn antenna is rotated by 90\degree{}.

\subsubsection{Long-Range Multi-Frequency Sounder}
The rest of the campaigns use a VNA-based sounder-setup capable of measurements in the 15-GHz, 28-GHz, 60-GHz, and 80-GHz bands and link distances up to about 200~m~\cite{Haneda2016, Naderpour2016, Nguyen2016Globecom , Vehmas2016, Virk2017}. 
In the 15-GHz sounder, the VNA is connected directly to the antennas using short radio frequency and long optical fiber cables, while a signal generator and frequency up and down converters are used to achieve the desired radio frequency at the 28-GHz, 60-GHz, and 80-GHz bands. 
The Tx and Rx antennas have identical radiation patterns across the studied frequency bands.  
On the Tx side, 2-dBi vertically-polarized omnidirectional biconical antenna with elevation beamwidth of 60\degree{} is used. 
For the main polarization measurement, the vertically-polarized Rx antenna is a directive $H$-plane sectoral horn antenna with a gain of 19 dBi and 10\degree{} and 40\degree{} beamwidths in the azimuth and elevation domains, respectively. 
For the cross-polarization measurement, the Rx antenna is an $E$-plane sectoral horn antenna that receives horizontally polarized fields mainly, with the same gain and beamwidths as the $H$-plane horn antenna.

\subsection{Power Angular Delay Profiles (PADPs)}
With both sounder setups, the directional channels were measured by rotating the horn in the azimuth plane from 0\degree{} to 360\degree{} with 5\degree{} steps\footnote{With a few exceptions: (i) in AIR, only a 140\degree{}-sector is recorded rather than the full 360\degree{}~\cite{Vehmas2016}, and (ii) in OFF1, OFF2, and STA at 71.5~GHz the horn was rotated by 1\degree{} steps and by 3\degree{} steps in C2 at 63~GHz.}, and a PADP is recorded. 
A Hamming window function is used in the inverse Fourier transform of the measured channel transfer functions in order to suppress the delay domain side-lobes. 
Maximum available measurement bandwidth (BW) has been used in all the measurements for the best possible delay resolution.\footnote{Maximum BW of the sounder setups are 4 and 5 GHz, but especially in outdoor environments radio link test license limits the allowed BW, see Table~\ref{table:Numbers}.} 
The VV-polarization measurement gives the main polarization PADP, i.e., $\mathrm{PADP}_{\mathrm{m}}(\tau,\varphi)$, and the VH-polarization measurement gives the cross-polarization PADP, i.e., $\mathrm{PADP}_{\mathrm{c}}(\tau,\varphi)$, where $\tau$ is the delay and $\varphi$ is the angle that the main-lobe of the horn antenna points towards.  

All the used antennas have high XPD levels. 
It is important to use high-XPD antennas in cross-polarization measurements as the antenna XPD limits the range of measurable XPR during wave propagation. 
Also, antenna tilt errors may cause the main polarization to leak to the cross-polarization measurement. 
The combined antenna polarization effects in the channel measurements can be observed from the polarization ratios of the direct paths in LOS links, as in~\cite{Karttunen2015}.


\subsection{Dual-Polarized Multipath Detection}
\label{subsec:MPCdetection}

The dual-polarized multipath detection aims to detect \textit{all detectable MPCs} and their main and the cross-polarization amplitudes. 
In terms of the MPC detection method, an important distinction needs to be made between channel models with only discrete paths, e.g.,~\cite{3GPP_TR38901,winner}, and models in which the channel is divided into the specular propagation paths and diffuse spectrum, e.g.,~\cite{Haneda2015, Haneda2014, Vehmas2016}. 

\begin{figure}[!t]
	\centering 
	\subfigure[]{
		\includegraphics[width=3.49in]{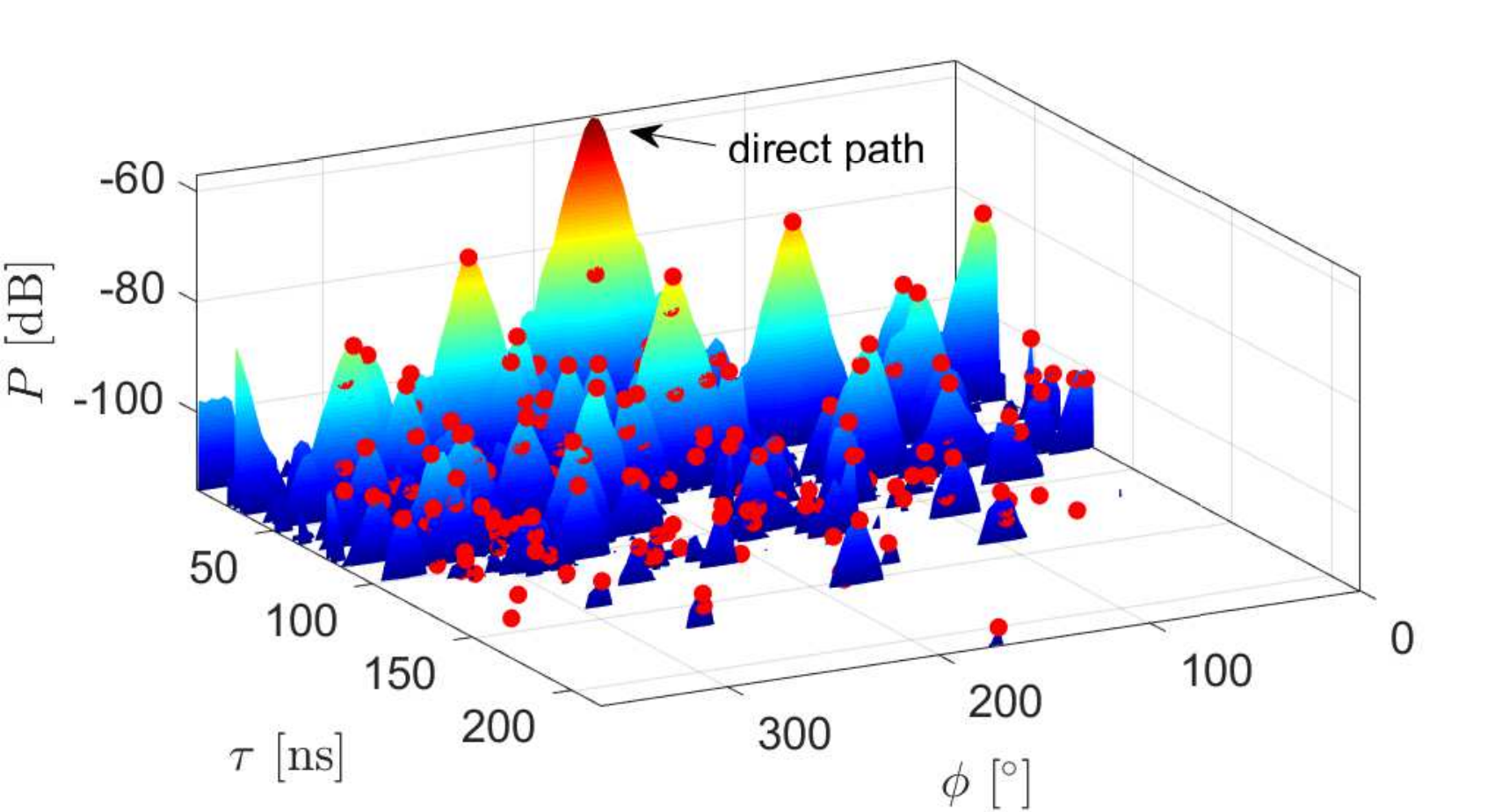}}
	\subfigure[]{
		\includegraphics[width=3.49in]{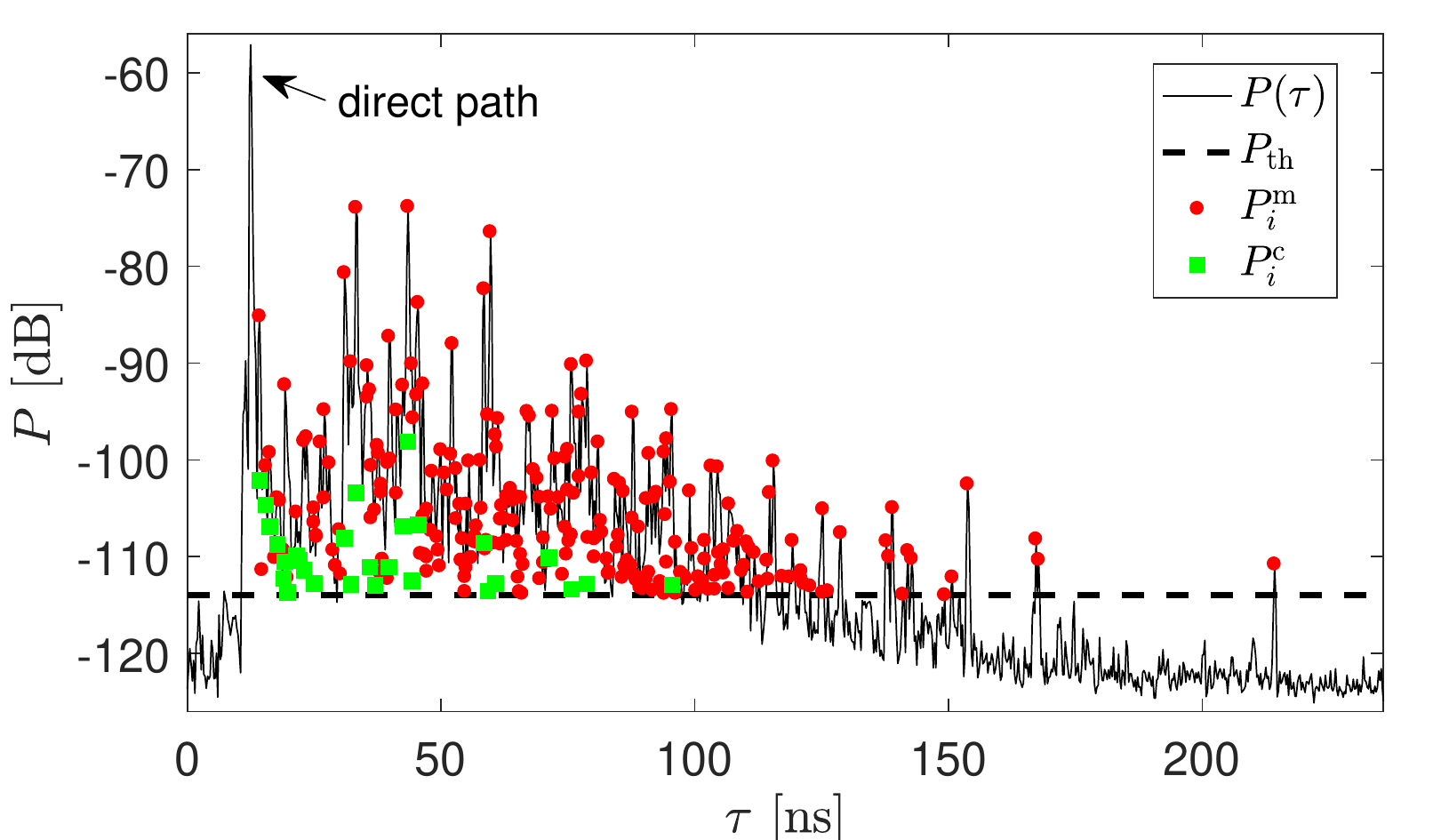}}
	\caption{Examples of (a) a measured PADP and (b) the delay domain representation $P(\tau)$ defined in~\eqref{eq:Ptau}. They are from 4-m long LOS link in cafeteria (C2) measured at 63~GHz. The markers represent detected local maximums. In total 246~paths are detected with 28~measured XPRs, 217~censured samples with $P_i^\mathrm{c}<P_{\mathrm{th}}$, and one censured sample with $P_i^\mathrm{m}<P_{\mathrm{th}}$.}
	\label{fig:padpexample_C2_n3}
\end{figure}

The amplitudes, delays, and angles are determined based on detected local maximums from the measured PADPs. 
The MPC detection is based on repeating two steps: 1) local maximum detection on the delay domain and 2) removal of the detected peaks from the PADP. 
These two steps are repeated until no more MPC can be found.
Both $\mathrm{PADP}_{\mathrm{m}}(\tau,\varphi)$ and $\mathrm{PADP}_{\mathrm{c}}(\tau,\varphi)$ are used in the multipath detection.\footnote{In~\cite{Karttunen2017}, only the main polarization PADP is used for MPC detection and step 1 is used only once resulting in fewer detected paths.}
A noise threshold level $P_{\mathrm{th}}$ is defined above the measurement noise level, as illustrated in Fig.~\ref{fig:padpexample_C2_n3}. 

The multipath detection consists of the following two steps.
\begin{enumerate}
	\item The delay domain representation of the PADP is defined as a maximum power over the angular domain as 
	\begin{equation}
	P(\tau)=\underset{\varphi}{\operatorname{max}} \:\{\mathrm{PADP}_{\mathrm{m}}(\tau,\varphi),\mathrm{PADP}_{\mathrm{c}}(\tau,\varphi)\}.
	\label{eq:Ptau}
	\end{equation}
	It must be noted that this is different from the PDP, as defined, e.g., in \cite{Haneda2015}, where the PDP is derived from the {\it mean} of the PADP over angles. 
	The amplitude of the i$^{\mathrm{th}}$ MPC $P_i$ and delay $\tau_i$ are detected as a local maximum of $P(\tau)$ with conditions
	\begin{equation}
	\label{eq:peak1}
	P_i(\tau_i) > \frac{1}{4\Delta_\tau}\int_{\tau-2\Delta_\tau}^{\tau+2\Delta_\tau}P(\tau)d\tau,
	\end{equation}
	\begin{equation}
	P_i(\tau_i) > P(\tau_i-\Delta_\tau), \quad
	P_i(\tau_i) > P(\tau_i+\Delta_\tau),
	\end{equation}
	where $\Delta_\tau$ is the delay resolution and $4\Delta_\tau$ is the length of a sliding window. 
	The condition \eqref{eq:peak1} ensures that only local maximums above a local average are detected as peaks in order to avoid detecting artifacts and noisy peaks. 
	The azimuth angle $\varphi_i$ is defined based on the maximum of the PADPs at delay $\tau_i$. 
	Additionally, $\tau_i$ needs to be greater than the delay of the direct path to avoid detecting the direct LOS path as an MPC.
	\item If any MPCs are detected in step 1 these MPCs are removed from the original PADPs before repeating step 1 to detect possible weaker MPCs than the previously detected peaks at the same delays.  
	An area of $\pm 6\Delta_\tau \times \pm 6\Delta_\varphi$ is removed from the PADPs; $\Delta_\varphi$ is the measurement angular resolution. 
	The size of the removed area needs to be large enough to cover the horn antenna main-beam. 
	These two steps are repeated until no more MPCs are detected.
\end{enumerate}

Having detected the MPC delays $\tau_i$ and angles $\varphi_i$, the main polarization $P_i^\mathrm{m}$ and cross-polarization $P_i^\mathrm{c}$ amplitudes are found from the PADPs. 
For each path $P_i$ is either $P_i^\mathrm{m}$ or $P_i^\mathrm{c}$ depending on which is stronger. 
The amplitude for the weaker is the maximum its PADP within $\pm 2\Delta_\tau \times \pm 2\Delta_\varphi$ of the MPC delay and angle. 
In this manner, we accept some uncertainty that the local maximums of the PADPs at the main and cross-polarization can appear at slightly different delays or angles. 
The peak detection includes some constants that are heuristically chosen to obtain meaningful local maximums for further analysis. 
These include the length sliding window $4\Delta_\tau$ in step 1), the area $\pm 6\Delta_\tau \times \pm 6\Delta_\varphi$ that is removed from the PADPs in step 2), and the tolerance $\pm 2\Delta_\tau \times \pm 2\Delta_\varphi$ that is allowed for the difference in the peaks main and cross-polarizations. 
An exemplary PADP overlaid with the detected MPCs are shown in Fig.~\ref{fig:padpexample_C2_n3}.

\subsection{Properties of the MPCs}

There are three types of detectable MPCs when it comes to the XPR and their detectability in channel sounding. Later these are called type 1, 2, and 3:
\begin{enumerate}
	\item Both the main and cross-polarization amplitudes of an MPC are above the noise threshold of the measurement, i.e., $P_i^\mathrm{m}>P_{\mathrm{th}}$ and $P_i^\mathrm{c}>P_{\mathrm{th}}$, and hence XPR~$=P_i^\mathrm{m}-P_i^\mathrm{c}$.
	\item Main polarization above and cross-polarization below the noise threshold, i.e., $P_i^\mathrm{m}>P_{\mathrm{th}}$ and $P_i^\mathrm{c}<P_{\mathrm{th}}$. In this case, XPR~$>P_i^\mathrm{m}-P_{\mathrm{th}}$.
	\item Cross-polarization above and main polarization below the noise threshold, i.e., $P_i^\mathrm{m}<P_{\mathrm{th}}$ and $P_i^\mathrm{c}>P_{\mathrm{th}}$. In this case,  XPR~$<P_{\mathrm{th}}-P_i^\mathrm{c}$.
\end{enumerate}

In general, the number of detectable MPCs depends on the measurement dynamic range, delay and angular resolution, a detection method of local maximums, link conditions including Tx-Rx separation distance, among others. 
The number of detected MPCs are listed in Table~\ref{table:Numbers} for the covered environments and frequency bands. 
Table~\ref{table:Numbers} also list the number of links and the dynamic range defined as $\max(P_i^\mathrm{m})-P_{\mathrm{th}}$, where $\max(P_i^\mathrm{m})$ is the strongest amplitude of the MPCs for the main polarization and $P_{\mathrm{th}}$ is the noise threshold level in that particular campaign.
It must be noted that $P_{\mathrm{th}}$ varies for radio frequencies because the measurement apparatus was not exactly the same.

From the 28 campaigns a total of 30862 MPCs are detected, 5611 are type~1 (with $P_i^\mathrm{m}>P_{\mathrm{th}}$ and $P_i^\mathrm{c}>P_{\mathrm{th}}$), 24674 are type~2 (with $P_i^\mathrm{m}<P_{\mathrm{th}}$), and 560 are type~3 (with $P_i^\mathrm{m}<P_{\mathrm{th}}$). 
The classification of the MPCs into the three detectable MPC types are also performed in each measurement campaign as summarized in Table~\ref{table:Numbers}. 
All the campaigns have more than 200~detected MPCs, out of which 14 have over 1000~MPCs. 
Especially the number of type~1 XPRs is an important measure of the resulting uncertainty of the XPR model. 
As three campaigns (SHOP and STA at 71.5~GHz, and SC3 at 14.25~GHz) have less than 50~MPCs with type~1 XPRs, they are not used to draw conclusions on the frequency or environment dependency of the XPR model in Section~\ref{secVI:XPRf}. 

\section{Parameter Estimation}
\label{secIV:parameterestimation}

Traditionally, the XPR model has been parameterized based on the measured XPR ratios, i.e., only type~1 MPCs with both main and cross-polarization above the noise threshold.
In fact, many detectable MPC have only one polarization component above the noise threshold. 
As shown in~\cite{Karttunen2017}, these so-called censored samples by the noise threshold of channel sounding can have a significant effect on the model parameter estimates. The censored samples can be taken into account using Tobit maximum likelihood estimation~\cite{Tobin1958, Karttunen2017, Gustafson2015, Karttunen2016IET}. 
In~\cite{Karttunen2017}, only type~1 and 2 MPCs were used. 
In this paper we use all detectable MPCs, including type 3 MPCs, to parameterize both the conventional and the improved XPR models defined as models~1 and 2 in Section~\ref{secII_Modeling_Depolarization}.

A log-likelihood function for the noise-censored XPR measurements is given by 
\begin{equation}
\begin{aligned}
L(\mu,\sigma)=\sum_{i=1}^{N} I_i  \left[-\ln(\sigma)+\ln\phi \left( \frac{P_i^\mathrm{m}-P_i^\mathrm{c}-\mu}{\sigma} \right)  \right] \\
+\sum_{i=1}^{N} (1-I_i)J_i \ln \left[1-\Phi \left( \frac{P_i^\mathrm{m}-P_{\rm{th}}-\mu}{\sigma} \right)  \right] \\
+\sum_{i=1}^{N} (1-I_i)(1-J_i)  \ln \left[\Phi \left( \frac{P_{\rm{th}}-P_i^\mathrm{c}-\mu}{\sigma} \right)  \right],
\end{aligned}
\label{eq:L}
\end{equation}
where $\mu$ and $\sigma$ are the mean and standard deviation of the log-normal XPR distribution \eqref{eq:XPR1} or \eqref{eq:XPR2}~-~\eqref{eq:mu2} and $N$ is the total number of detected MPC in a campaign. 
The three detectable types of MPC are indicated with indexes $I$ and $J$; type~1 with $I_i=1$, type~2 with $I_i=0$ and $J_i=1$, and type~3 with $I_i=0$ and $J_i=0$. 
Functions $\phi(\cdot)$ and $\Phi(\cdot)$ are the probability density function (PDF) and cumulative density function (CDF) of the normal distribution. 

The parameter estimates are derived as
\begin{equation}
[\hat{\mu}_1,\hat{\sigma}_1] = \underset{}{\operatorname{arg}} \underset{\mu_1,\sigma_1}{\operatorname{min}} \{-L(\mu_1,\sigma_1)\},
\label{eq:argmin_1}
\end{equation}
\begin{equation}
[\hat{\alpha}_2,\hat{\beta}_2,\hat{\sigma}_2] = \underset{}{\operatorname{arg}} \underset{\alpha_2,\beta_2,\sigma_2}{\operatorname{min}} \{-L(\mu_2(L_\mathrm{ex}),\sigma_2)\},
\label{eq:argmin_2}
\end{equation}
where $\hat{\cdot}$ denotes parameter estimates and $\alpha_2$ and $\beta_2$ are the parameters in the excess loss dependent function \eqref{eq:mu2}.

\begin{figure}[!t]
	\centering 
	\includegraphics[width=3.5in]{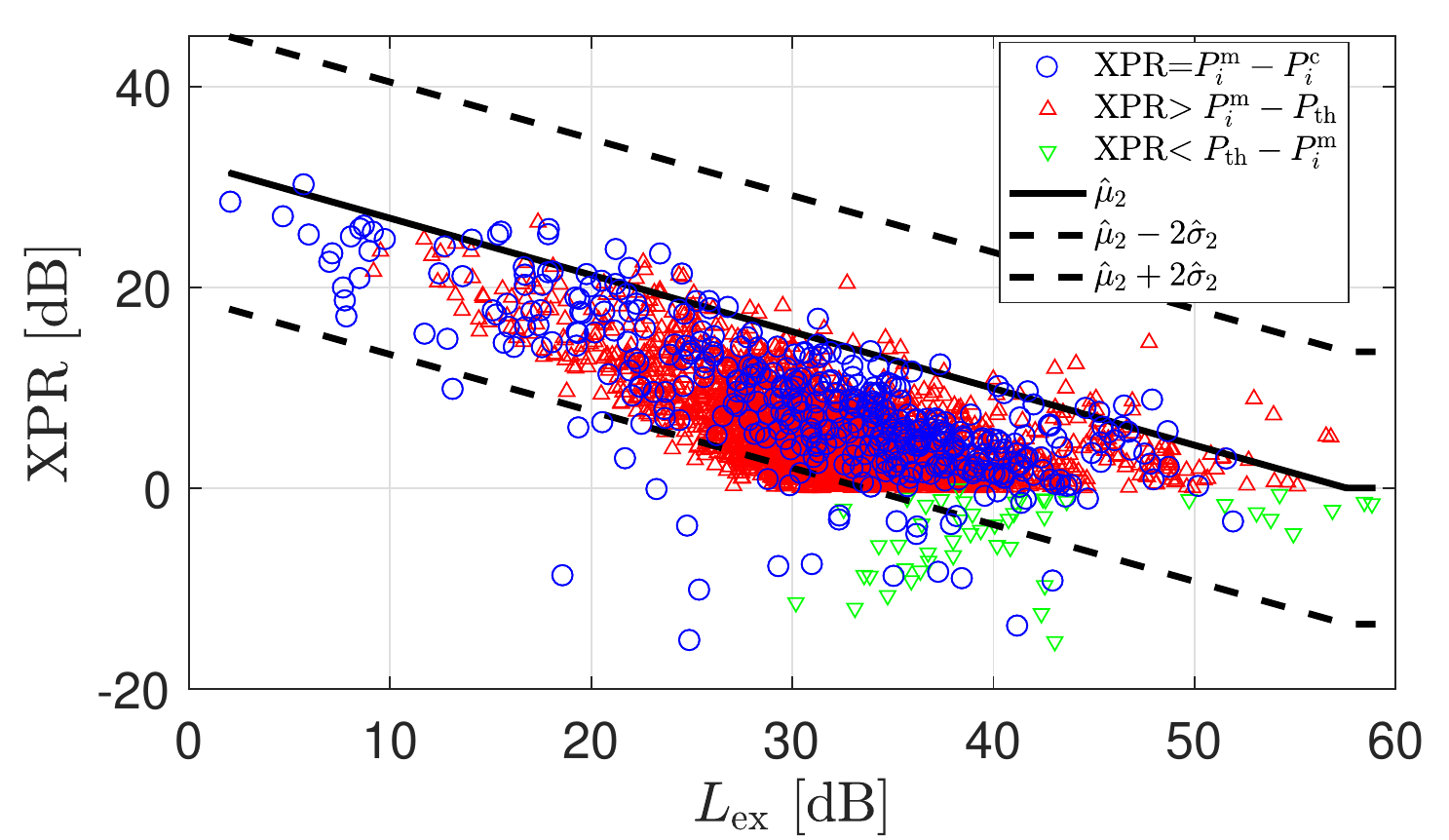}
	\caption{Measured XPRs (XPR $=P_i^\mathrm{m}-P_i^\mathrm{c}$), censored samples (XPR $>P_i^\mathrm{m}-P_{\mathrm{th}}$ and XPR $<P_{\mathrm{th}}-P_i^\mathrm{c}$), and fitted model $\hat{\mu}_2$ (solid line) and $\hat{\mu}_2\pm 2\hat{\sigma}_2$-limits (dash lines) in open square (SQR2) at~61~GHz. The number of detectable MPCs are 422, 2439, and 48, for type~1, 2, and 3, respectably. The fitted model~2 is $\hat{\mu}_2=-0.57\cdot L_{\mathrm{ex}}+33$ and $\hat{\sigma}_2$~=~6.8.}
	\label{fig:XPRvsLex_SQR2}
\end{figure}

An example of MPC XPR measurements, and their fitting with the excess loss dependent model, as a function of the excess loss, is shown in  Fig.~\ref{fig:XPRvsLex_SQR2}. 
The parameter estimates for each campaign and for both models are listed in Table~\ref{table:Numbers}. 
It was found that the MPC XPR is always high for low $L_\mathrm{ex}$-paths (large $\hat{\beta}_2$) and inversely proportional to $L_\mathrm{ex}$ (negative $\hat{\alpha}_2$). 
As in~\cite{Karttunen2017}, we can see that the model~2 fits the data better than the model~1 as the standard deviation $\hat{\sigma}_2$ is much smaller than $\hat{\sigma}_1$.\footnote{In fact, since model~2 has more optimized parameters, $\hat{\sigma}_2$ is always smaller than $\hat{\sigma}_1$ (except if $\hat{\alpha}_2=0$, then $\hat{\beta}_2=\hat{\mu}_1$ and $\hat{\sigma}_2=\hat{\sigma}_1$).}
The clearly observed strong dependence of the MPC XPR on the excess loss motivates the use of model~2.

\section{Model Comparison}
\label{secV:XPRmodelComparison}

In this section, the improved MPC XPR model is compared with the conventional one based on the accuracy of reproducing the measurements in terms of the total cross-polarization power. 
The aim is to derive a single-number accuracy metric that can be used to determine which MPC XPR model is better. 
The comparison of the standard deviations $\hat{\sigma}_1$ and $\hat{\sigma}_2$ in Table~\ref{table:Numbers} clearly shows that the model~2 fits the data better on an MPC-level. 
It is important to use a different metric for model validation than what is used for the parametrization. 
The prediction accuracy of total cross-polarization power can be seen as an \textit{indirect validation method} in which more importance is given to the more important strong paths. 

The comparison is made between the measured and synthesized cross-polarization powers. 
The synthesized total cross-polarization is calculated based on the \textit{measured main polarization} and the MPC XPR models.
The total measured cross-polarization power $C_{\mathrm{tot}}$ is calculated as a sum of the above noise threshold MPC cross-polarization amplitudes $P_i^\mathrm{c}$ as
\begin{equation}
C_{\mathrm{tot}}= 10\log_{10} \left( \sum_{i=1}^{L^{\mathrm{c}}} 10^{P_i^\mathrm{c}/10} \right),
\label{eq:Ctot}
\end{equation}
where $L^{\mathrm{c}}$ is the number of MPCs with detectable cross-polarization level in a link. 
 $C_{\mathrm{tot}}$ is calculated for each link. 
Occasionally $P_i^\mathrm{c}<P_{\mathrm{th}}$ for all MPCs and the total power is censored $C_{\mathrm{tot}}<P_{\mathrm{th}}$.

The synthesized cross-polarization power for $i^{\mathrm{th}}$ MPC is
\begin{equation}
\tilde{P}_i^\mathrm{c} = \begin{cases} P_i^\mathrm{m}-XPR, & \mbox{if } P_i^\mathrm{m}>P_\mathrm{th} \\ P_i^\mathrm{c}, & \mbox{if } P_i^\mathrm{m}<P_\mathrm{th}, \end{cases}
\end{equation}
where $XPR$ is independent random XPR-value drawn from one of the MPC XPR models and  $P_i^\mathrm{m}$ and $P_i^\mathrm{c}$ are the measured main and cross-polarization powers in dB's. 
The same noise threshold as measurements, $P_{\mathrm{th}}$, is used when reproducing censored MPCs.  
The total synthesized cross-polarization power $\tilde{C}_{\mathrm{tot}}$ is then
\begin{equation}
\tilde{C}_{\mathrm{tot}}= 10\log_{10} \left( \sum_{i=1}^{\tilde{L}} 10^{\tilde{P}_i^\mathrm{c}/10} \right),
\label{eq:synthCtot}
\end{equation}
where $\tilde{L}$ is the number synthesized cross-polarization components above the noise threshold. 
If all $\tilde{P}_i^\mathrm{c}<P_{\mathrm{th}}$ then $\tilde{C}_{\mathrm{tot}}$ is censored ($\tilde{C}_{\mathrm{tot}}<P_{\mathrm{th}}$) similarly as with the measured values. 
Since $\tilde{C}_{\mathrm{tot}}$ is censored data, the Tobit maximum likelihood estimation can be used to estimate the mean and standard deviation of $\tilde{C}_{\mathrm{tot}}$.

The difference between the synthesized and the measured total cross-polarizations, in dB's, is
$\epsilon= \tilde{C}_{\mathrm{tot}}-C_{\mathrm{tot}}$.
It is possible to assess the ability of the MPC XPR model to reproduce the measurements for each radio frequency and measurement site by estimating $\epsilon$ with many $\tilde{C}_{\mathrm{tot}}$ in a Tx-Rx link, and then taking a mean of the estimates. 
Since a part of estimates, $\tilde{C}_{\mathrm{tot}}$ and $C_{\mathrm{tot}}$ are censored by the noise, the model error term $\epsilon$ is also subject to the censoring. 
Similarly as the MPC XPR estimates there are three different estimates of the error term: a uniquely determined $\epsilon$ values, $\epsilon< P_{\mathrm{th}}-C_{\mathrm{tot}}$, and $\epsilon> \tilde{C}_{\mathrm{tot}}-P_{\mathrm{th}}$. 
The Tobit maximum likelihood can be used to estimate the mean and standard deviation of the $\epsilon$'s.  
The mean values $\mu_\epsilon$ are given in Table~\ref{table:Numbers}. 
A good model gives $\mu_\epsilon\approx 0$~dB. 
A positive $\mu_\epsilon$ means that the model typically overestimates the total cross-polarization power for the measured MPCs. 
As $\epsilon$ is defined based on the total cross-polarization power it mainly measures the prediction accuracy for the highest cross-polarization amplitudes generally associated with the strongest MPCs. 
A comparison of the conventional and the improved XPR models based on $\epsilon$ reveals that the model~1, with constant $\mu_1$ and $\sigma_1$, does not reproduce the measured total cross-polarization accurately and overestimates it by about 4 to 10~dB. 
The model~2 has lower $\mu_\epsilon$ in every environment and frequency with typical values between -1 and 4~dB. 
The better accuracy of total cross-polarization power modeling is mostly due to the right modeling of the XPR for the strongest MPCs.

\section{Frequency Dependency}
\label{secVI:XPRf}

\begin{figure}[t!]
	\centering
	\includegraphics[width=3.133in]{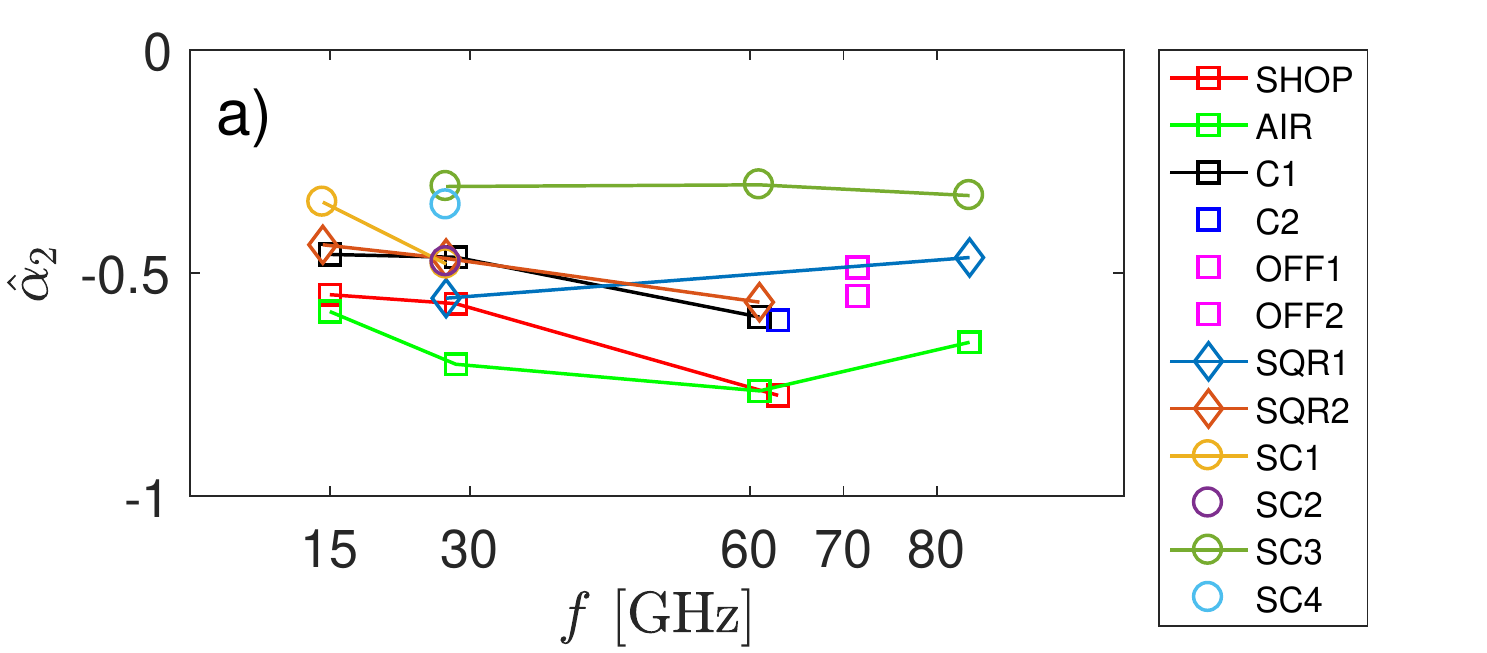}
	\label{fig:xpr_a2}
	\includegraphics[width=3.133in]{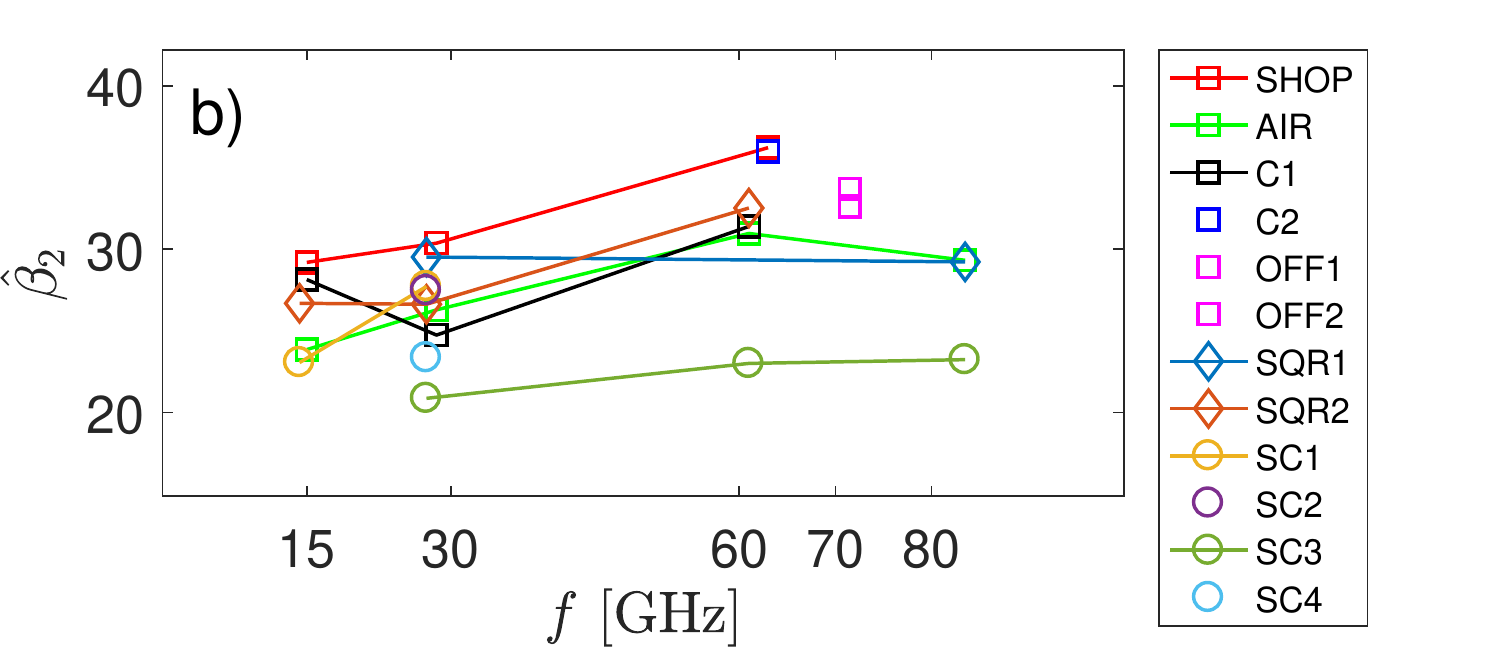}
	\label{fig:xpr_b2}
	\includegraphics[width=3.133in]{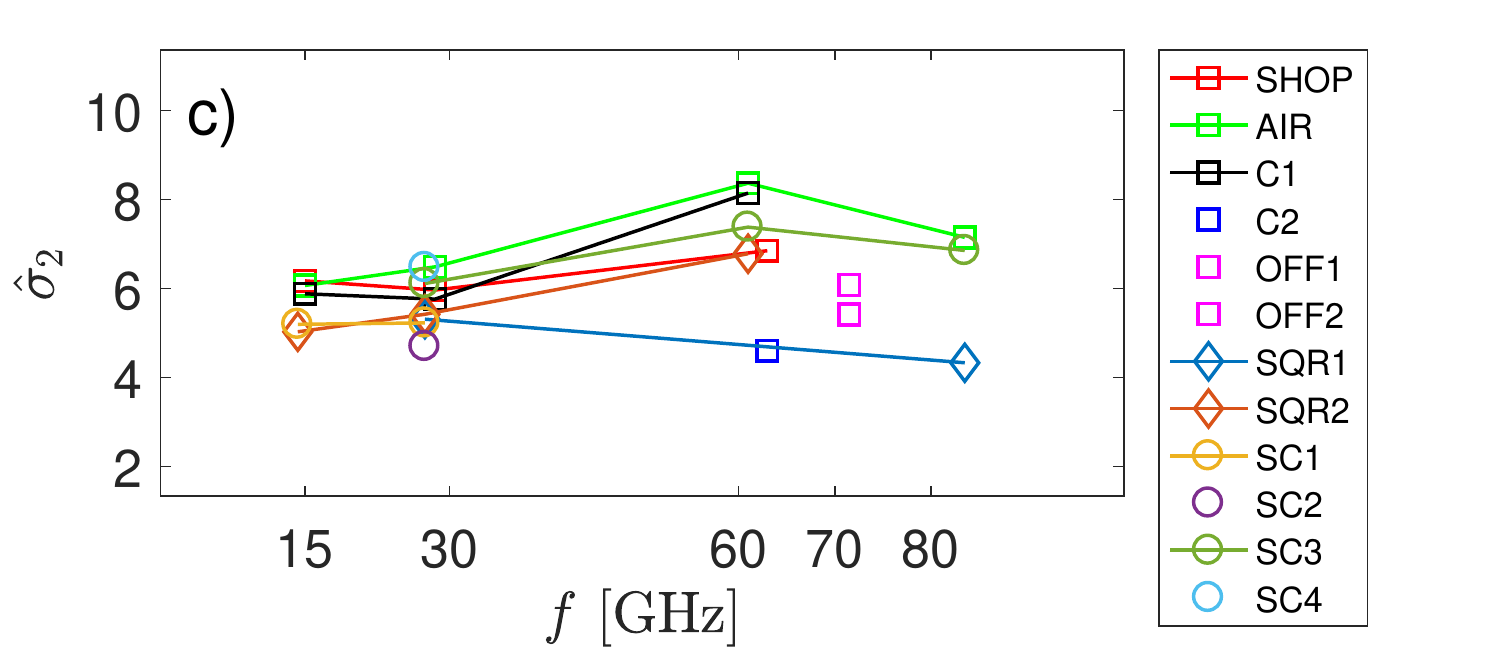}
	\label{fig:xpr_s2}
	\caption{Parameter estimates a) $\hat{\alpha}_2$, b) $\hat{\beta}_2$, and c) $\hat{\sigma}_2$ as a function of the frequency for various indoor (SHOP, AIR, C1, and C2) and outdoor (SQRs and SCs) environments.}
	\label{fig:xpr2_estimats_vs_f}
\end{figure}

\begin{table}[t!]
	\caption{Average Parameter Estimates.}
	\label{table:mean}
	\centering
	\begin{tabular}{c||ccc|}
		& $\alpha_2$ & $\beta_2$ & $\sigma_2$ \\
		\hline
		Average &-0.5&28&6\\
	\end{tabular}
\end{table}

The parameter estimates of the improved XPR model (model~2) are presented in Fig.~\ref{fig:xpr2_estimats_vs_f} against the center frequency of channel sounding. 
In this section, the three measurement campaigns with relatively small sample size have been omitted from the analysis. 
The estimated parameters show no clear frequency dependency. 
The parameters are only slightly location dependent and no clear difference can be observed between indoor and outdoor environments.
Therefore, we propose to use average values shown in Table~\ref{table:mean} for the covered radio frequencies and environments in our channel sounding.
The proposed MPC XPR model can be summarized as (\ref{eq:XPR2})-(\ref{eq:mu2}) with parameter values from Table~\ref{table:mean}.

It should be noted that the observed frequency and environment independence of the model parameters do not imply that the channel polarization properties are independent of the frequency and environment. 
The statistical distribution of the MPC excess losses is likely dependent on the frequency and environment resulting in differences in the channel polarization properties in the link level. 

\section{Conclusion}
\label{conclusion}

The frequency and environment dependency of MPC XPR are studied based on 28 measurement campaigns in both indoor and outdoor environments in the cm-wave (15-GHz, 28-GHz) and mm-wave (60-GHz, 70-GHz, and 80-GHz) bands. 
The campaigns include a total of 265 links and 30862 detected MPCs. 

Two models of MPC XPR based on log-normal distribution are parameterized and compared. 
It is found from our measurements that a mean XPR in dB-scale linearly decreases as the MPC excess loss increases. 
The finding leads us to the improved MPC XPR model for above 6~GHz, in which zero-dB excess-loss paths have a mean XPR of 28~dB and the mean decreases half-a-dB for every dB of MPC excess loss; the standard deviation around the mean is 6~dB. 
The improved model intuitively makes physical sense when it comes to interaction between propagating waves and physical objects. 
It was demonstrated that the model outperforms a conventional MPC XPR model with a constant mean in terms of fitting accuracy. The validity of the new model is strengthened by the analysis of the total cross-polarization power. Our study finally revealed that the MPC XPR shows no clear frequency or environment dependency. 

The model is usable in the existing geometry-based stochastic channel models instead of the typical MPC XPR model with the constant mean value.
In future, we aim to examine the difference between the below 6~GHz and the above 6~GHz frequency ranges.

\section*{Acknowledgment}
We would like to thank Afroza Khatun, Mikko Kyr\"{o}, Lauri Laakso, Niko Lindvall, Reza Naderpour, Pekka Rummukainen, Joni Vehmas, Miao Yang, and Usman Virk for help during the measurements.



\end{document}